\documentclass[floatfix,reprint,amsmath,amssymb,aps,pra]{revtex4-1}
\usepackage{times,mathptmx}
\usepackage{graphicx}
\usepackage[english]{babel}
\usepackage{epstopdf}
\usepackage{xcolor}

\begin{document}

\title{Controlling wave-front shape and propagation time with tunable disordered non-Hermitian multilayers}

\author{Denis~V.~Novitsky$^{1}$}
\email{dvnovitsky@gmail.com}
\author{Dmitry~Lyakhov$^{2}$}
\author{Dominik~Michels$^{2}$}
\author{Dmitrii Redka$^3$}
\author{Alexander~A.~Pavlov$^{4}$}
\author{Alexander~S.~Shalin$^{5}$}

\affiliation{$^1$B.~I.~Stepanov Institute of Physics, National
Academy of Sciences of Belarus, Nezavisimosti Avenue 68, 220072
Minsk, Belarus \\ $^2$Visual Computing Center, King Abdullah University of Science and Technology, Thuwal 23955-6900, Kingdom of Saudi Arabia\\ $^3$ Saint Petersburg, Electrotechnical University “LETI” (ETU),  Prof. Popova Street 5, 197376 St. Petersburg, Russia \\ $^4$ Institute of Nanotechnology of Microelectronics of the Russian Academy of Sciences, Leninsky Prospekt 32A, 119991 Moscow, Russia \\ $^5$ITMO University, Kronverksky Prospekt 49,
197101 St. Petersburg, Russia}

\date{\today}

\begin{abstract}
Unique and flexible properties of non-Hermitian photonic systems attract ever-increasing attention via delivering a whole bunch of novel optical effects and allowing for efficient tuning light-matter interactions on nano- and microscales. Together with an increasing demand for the fast and spatially compact methods of light governing, this peculiar approach paves a broad avenue to novel optical applications. Here, unifying the approaches of disordered metamaterials and non-Hermitian photonics, we propose a conceptually new and simple architecture driven by disordered loss-gain multilayers and, therefore, providing a powerful tool to control both the passage time and the wave-front shape of incident light with different switching times. For the first time we show the possibility to switch on and off kink formation by changing the level of disorder in the case of adiabatically raising wave fronts. At the same time, we deliver flexible tuning of the output intensity by using the nonlinear effect of loss and gain saturation. Since the disorder strength in our system can be conveniently controlled with the power of the external pump, our approach can be considered as a basis for different active photonic devices.
\end{abstract}

\maketitle

\section{Introduction}\label{intro}

Recently, the studies of open optical systems containing loss and gain attract increased attention. Although such systems are well-known for many years, the recent trend of non-Hermitian photonics provides the second breath to the investigations of lasers, waveguides, resonators, etc. This is not only due to a different language borrowed from quantum mechanics, but also because of a number of novel phenomena found in loss-gain structures. We name here only a few examples, such as the effects of $\mathcal{PT}$ symmetry \cite{Zyablovsky2014, Feng2017, El-Ganainy2018} and exceptional points \cite{Ozdemir2019, Miri2019}. These effects include unidirectional invisibility \cite{Lin2011, Feng2013}, sensors \cite{Chen2017,Hodaei2017} and gyroscopes \cite{Hokmabadi2019, Lai2019} with enhanced sensitivity, loss-induced \cite{Peng2014} and asymmetric \cite{Novitsky2018} lasing, novel single-mode \cite{Feng2014-2, Hodaei2014, Gu2016} and vortex \cite{Zhang2020} lasers, coherent perfect absorbers \cite{Longhi2010, Chong2011, Wong2016, Novitsky2019a}, and topological edge-boundary correspondence \cite{Leykam2017, Takata2018, Ni2018, Zhao2019}.

Disordered photonics is another spotlight of modern research \cite{Wiersma2013}. It deals with light propagation in the presence of random fluctuations of the medium parameters such as refractive index or unit cell dimensions. The rich physics of such systems rooted in multiple scattering allows to realize a number of unusual features; the scattering properties of single particles and complex structures are well-studied in literature \cite{Zhigunov2018, Baryshnikova2018, Ivinskaya2018}. The most prominent feature of disordered systems is the Anderson localization of light \cite{Segev2013, Gredeskul2012, Sheinfux2017} appearing as a result of multi-path interference of waves scattering on random inclusions. Multiple scattering can also lead to the peculiar statistical properties of light violating usual diffusion (sub- and superdiffusion) like in optical Levy flights \cite{Chabanov2003, Sarma2014, Naraghi2016}. The situation becomes even more complicated when the interplay between disorder and nonlinearity occurs with the subsequent suppression of Anderson localization or promotion of diffusion \cite{Conti2007, Lahini2008, Sharabi2018}.

There is a recent trend combining together disorder and non-Hermiticity in the random systems with loss and gain. One of the main aims of such combinations is the enhancement of transmission which is usually strongly suppressed due to multiple scattering \cite{Makris2015}. For example, this problem can be solved with the help of the concept of so-called constant-intensity waves in specially designed loss-gain profiles \cite{Makris2015, Makris2017, Brandstotter2019, Tzortzakakis2020, Makris2020}. The non-Hermitian disorder due to random fluctuations of loss and gain can be a source of novel-type localized states \cite{Hamazaki2019, Tzortzakakis2020a, Huang2020}. There is also a very active subfield of random lasing obtained in disordered amplifying media \cite{Noginov2005, Turitsyn2010, Liu2014, Abaie2017}. However, the \textit{dynamics} of light interaction with structures containing both loss/gain and disorder are still poorly studied.

In this paper, we analyze the propagation of wave fronts through the disordered loss-gain non-Hermitian multilayer structures. The problem is aimed to be as realistic as possible: The fronts are the monochromatic waveforms having finite switching time, the loss and gain are due to resonant media and not merely a phenomenological imaginary part of permittivity, and the disorder can be controlled with external pump and change in time due to gain depletion and loss saturation. We have previously reported the study of short pulse propagation and localization in such media \cite{Novitsky2019} with the possibility to slow down or even stop the pulse. Here, we deal with the opposite case of continuous radiation with the emphasis on the transient process of steady-state establishment for the light intensities large enough to saturate the medium and give substantial transmission. This process can have different dynamics depending on the sharpness of the incident wave switching. In particular, we distinguish two regimes, when switching is slow (adiabatic) and fast (non-adiabatic). For these switching regimes, we show that the introduction of disorder changes the characteristic time of the transient process, whereas the resulting intensity of the signal is governed by saturation-limited input intensity. In particular, for the first time we demonstrate how the disorder can be used to switch on and off kinks at the output of the system. Thus, the non-Hermitian approach to disorder-induced control of propagation time, wave-front shape and transmitted intensity proposed in this paper opens new possibilities for ultrafast (picosecond or subnanosecond) multifunctional manipulation of optical signals.

\section{Problem statement}\label{eqs}

Hereinafter, we consider a host dielectric doped with two-level atoms. Light propagation in such a medium is described by
the well-known semiclassical Maxwell-Bloch equations for the dimensionless electric-field amplitude
$\Omega=(\mu/\hbar \omega) E$ (normalized Rabi frequency), complex
amplitude of the atomic polarization $\rho$, and population
difference between the ground and excited state $w$
\cite{Allen,Crenshaw2008,Novitsky2011}:
\begin{eqnarray}
\frac{d\rho}{d\tau}&=& i l \Omega w + i \rho \delta - \gamma_2 \rho, \label{dPdtau} \\
\frac{dw}{d\tau}&=&2 i (l^* \Omega^* \rho - \rho^* l \Omega) -
\gamma_1 (w-1),
\label{dNdtau} \\
\frac{\partial^2 \Omega}{\partial \xi^2}&-& n_d^2 \frac{\partial^2
\Omega}{\partial \tau^2}+2 i \frac{\partial \Omega}{\partial \xi}+2
i n_d^2 \frac{\partial \Omega}{\partial
\tau} + (n_d^2-1) \Omega \nonumber \\
&&=3 \epsilon l \left(\frac{\partial^2 \rho}{\partial \tau^2}-2 i
\frac{\partial \rho}{\partial \tau}-\rho\right), \label{Maxdl}
\end{eqnarray}
where $\tau=\omega t$ and $\xi=kz$ are the dimensionless time and
distance, $\mu$ is the dipole moment of the quantum transition,
$\hbar$ is the reduced Planck constant, $\delta=\Delta
\omega/\omega=(\omega_0-\omega)/\omega$ is the normalized frequency
detuning, $\omega$ is the carrier frequency, $\omega_0$ is the
frequency of the quantum transition, $\gamma_{1}=1/(\omega T_{1})$
and $\gamma_{2}=1/(\omega T_{2})$ are the normalized relaxation
rates of population and polarization respectively, and $T_1$ ($T_2$)
is the longitudinal (transverse) relaxation time; $\epsilon= \omega_L / \omega = 4 \pi \mu^2 C/3 \hbar
\omega$ is the light-matter coupling strength with $C$
the density of two-level atoms and $\omega_L$ the Lorentz frequency; $l=(n_d^2+2)/3$ is the local-field
enhancement factor due to the polarization of the host
dielectric with refractive index $n_d$ by the embedded two-level
particles. We numerically solve Eqs.
(\ref{dPdtau})--(\ref{Maxdl}) using the finite-difference approach
described in Refs. \cite{Crenshaw1996,Novitsky2009} well-proven in solving such tasks.

The parameters used for calculations are characteristic, e.g., for semiconductor quantum dots as the active particles. We suppose the exact resonance ($\delta=0$). The host refractive index is $n_d=1.5$. Incident monotonically switching cw field has the central wavelength $\lambda=0.8$ $\mu$m and the envelope as follows,
\begin{eqnarray}
\Omega(t)= \frac{\Omega_0}{1+e^{-(t-t_0)/t_p}}, \label{waveform}
\end{eqnarray}
where $\Omega_0$ is the amplitude of the resulting cw field (plateau), $t_p$ is the switching time, $t_0=5t_p$ is the offset time.

\begin{figure}[t!]
\centering \includegraphics[scale=1, clip=]{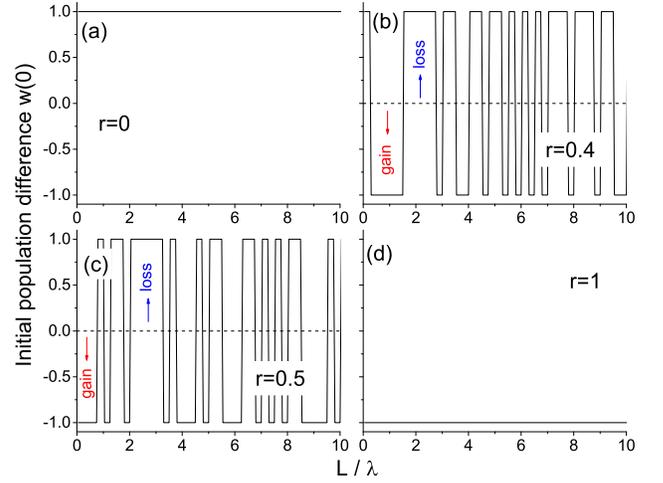}
\caption{\label{fig0} The example of initial population difference distributions for different disorder strengths $r$.}
\end{figure}

The disorder is introduced to the system through the periodical random variations of the initial population difference $w_0=w(t=0)$ along the light propagation direction. We use here the two-valued quadratic model of disorder described in Ref. \cite{Novitsky2019}. In fact, we have the multilayer structure with the initial population difference in the $j$th
layer of the medium corresponding to the distance $(j-1) \delta L < z
\leq j \delta L$ given by
\begin{eqnarray}
w^{(j)}_0 = \textrm{sgn} (1 - 2 r [(2\zeta_j-1)(r-1)+1]), \label{randvar}
\end{eqnarray}
where $\zeta_j$ is the random number uniformly distributed in the
range $[0; 1]$, $r$ is the parameter of the disorder strength, $\textrm{sgn}$ is the sign function, and $\delta L=\lambda/4$ is the layer thickness. When $r=0$, we have the trivial case of purely absorbing medium
[all $w^{(j)}_0=1$, Fig. \ref{fig0}(a)]. For $r \gtrsim 0.3$, the gain layers with $w^{(j)}_0=-1$ become possible [Figs. \ref{fig0}(b) and \ref{fig0}(c)]. The case of the maximal disorder, $r=1$, corresponds to the purely amplifying medium (all $w^{(j)}_0=-1$, Fig. \ref{fig0}(d)]. Thus, the parameter $r$ not only governs deviation from the ordered case of pure loss, but also takes on the role of pumping strength resulting
in appearance of gain. In general, the different layers of the structure are under different, randomly distributed pumping and can be lossy or gainy with a certain probability. This can be realized in a side-pumping scheme similar to that utilized in Ref. \cite{Wong2016} or with the adaptive-pumping approach \cite{Bachelard2014}. Note that the quadratic model of disorder (\ref{randvar}) give essentially the same results as the linear one \cite{Novitsky2019}, but is more convenient for symmetric representation of gain and loss. The similar linear model was experimentally realized recently in the context of random lasing \cite{Lee2019}.

\section{Non-adiabatic fronts}\label{nadiab}

In this section, we consider the case of non-adiabatically switching field (\ref{waveform}), when $t_p \ll T_2$. In particular, we take the relaxation times $T_1=1$ ns and $T_2=0.1$ ns and the switching time $t_p=5$ ps. The Lorentz frequency is $\omega^0_L = 10^{10}$ s$^{-1}$. The full thickness of the medium is $L=100 \lambda$. The final amplitude is $\Omega_0=10 \gamma_2$.

\begin{figure}[t!]
\centering \includegraphics[scale=1, clip=]{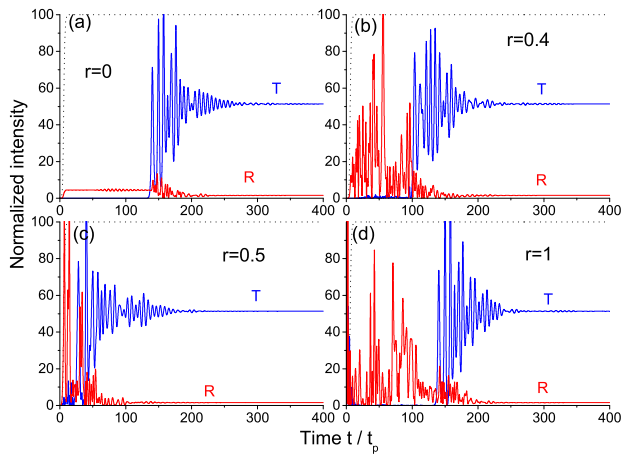}
\caption{\label{fig1} Intensity profiles for the transmitted and reflected light in the case of incident \textit{non-adiabatic} front. Different panels show the results for different disorder strengths $r$.}
\end{figure}

Figure \ref{fig1} shows the results of transmitted and reflected intensities calculations for different values of the disorder strength $r$. The initial population difference used in calculations is the same as in Fig. \ref{fig0}. Note that we consider here a single realization of disorder, since the observed features of interest for us are the same for different realizations at a certain level of disorder $r$: the specific oscillations of output intensity can differ, but the time needed for steady-state establishment and the final intensity are essentially the same for every realization. In the ordered system [$r=0$, see Fig. \ref{fig1}(a)], which is the uniform resonantly absorbing medium, the time needed for the front to pass through ($\sim 150 t_p$) is much longer due to dispersion than the free propagation time ($L n_d/c \sim 0.08 t_p$) and, in fact, is governed by the relaxation time $T_2=20 t_p$. The stationary response of the medium is established after its saturation and is seen from $\sim 300 t_p$ on.

Increasing disorder strength $r$ results in larger number of gain layers. As a result of stimulated emission in these layers, saturation needs less time and the stationary level is established faster. This is especially obvious for $r=0.5$ [Fig. \ref{fig1}(c)]. For even larger $r$, the number of gain layers becomes so large that amplified emission in the form of powerful bursts happens in the very first instants of time effectively returning most two-level particles to the ground level. Therefore, the propagation time of the front through the highly-amplifying system is similar to that in the case of purely absorbing medium [compare Figs. \ref{fig1}(d) and \ref{fig1}(a)].

\begin{figure}[t!]
\centering \includegraphics[scale=1, clip=]{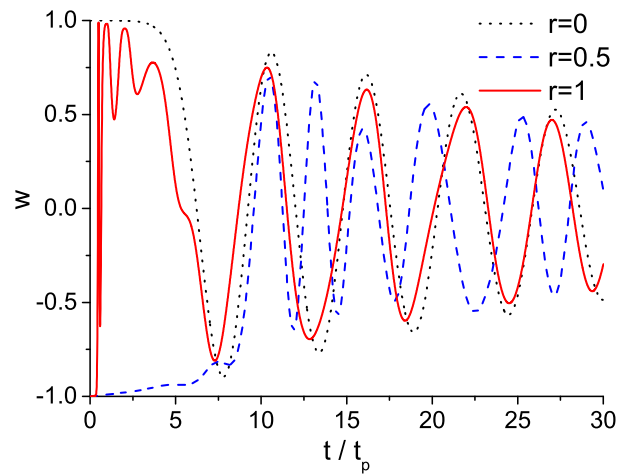}
\caption{\label{fig2} Dynamics of population difference at the entrance of non-adiabatic front in the medium with different disorder strengths $r$.}
\end{figure}

The difference in the response time can be illustrated with the dynamics of population difference at the entrance of the structure shown in Fig. \ref{fig2}. This figure demonstrates the initial stage of medium saturation with the oscillations converging to the very low (almost zero) value. These are the well-known Rabi oscillations with the frequency given by the so-called Rabi frequency and, hence, dependent on the incident radiation amplitude. Note that for these oscillations to appear, the Rabi frequency should be larger than the medium relaxation rate that is easily satisfied in our calculations ($\Omega_0=10 \gamma_2$). It is seen that although the dynamics for the amplifying ($r=1$) and the absorbing medium ($r=0$) start from absolutely different levels ($w=-1$ and $w=1$, respectively), the oscillations of population difference very closely follow each other. On the contrary, for $r=0.5$, we also start from $w=-1$ (the first layer with gain), but the subsequent dynamics strongly differs from those for $r=0$ and $r=1$. This confirms that similarity of the transmitted intensity profiles in Figs. \ref{fig1}(d) and \ref{fig1}(a) is not accidental.

\begin{figure}[t!]
\centering \includegraphics[scale=1, clip=]{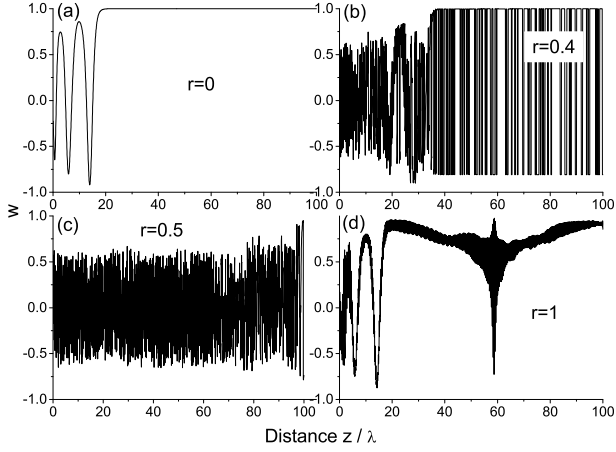}
\caption{\label{fig3} Distributions of population difference along the medium with different disorder strengths $r$ at the time instant $t=20 t_p$. The medium is excited by the non-adiabatic wave front.}
\end{figure}

An additional corroboration of this conclusion is given in Fig. \ref{fig3}, which shows the distribution of population difference for different disorder strengths $r$ at the time instant $t=20 t_p$ corresponding to the initial stage of radiation interaction with the medium (before the steady-state is established). It is seen that we have the chaotic distributions at $r=0.4$ [Fig. \ref{fig3}(b)] and $r=0.5$ [Fig. \ref{fig3}(c)] which can be treated as the variations of population difference around zero value. In other words, the medium can be considered as saturated on average. The light-matter interaction is comparatively weak in this case (there is no loss and gain on average) resulting in the increased speed of signal propagation, especially for $r=0.5$. On the contrary, the distributions for the purely absorbing [r=0, Fig. \ref{fig3}(a)] and purely amplifying media [r=1, Fig. \ref{fig3}(d)] are very similar, except for some local excitation due to random wanderings of light inside the medium. This confirms the rapid relaxation of amplifying medium due to spontaneous emission, so that the incident wave front propagates further in such effectively de-inverted medium giving the response analogous to that for $r=0$.

We see from Fig. \ref{fig1} that the stationary level of transmission is around $50 \%$ and reflection is only a few $\%$. The rest (almost half the energy of the wave) is absorbed by the saturated medium. How realistic is it? Let us estimate the level of stationary population difference necessary for this value of absorption. In the steady-state approximation, the two-level medium can be described with the effective dielectric permittivity as follows \cite{Novitsky2017}
\begin{eqnarray}
\varepsilon_{eff} = \varepsilon'_{eff} + \varepsilon''_{eff} = n_d^2 + \frac{K (-\delta + i \gamma_2)}{1 + |\Omega|^2/\Omega_{sat}^2}, \label{effeps}
\end{eqnarray}
where $\Omega_{sat}^2 = \gamma_1 (\gamma_2^2 + \delta^2)/4 l^2 \gamma_2$ is the saturation intensity, $K=3 \omega_L l^2/\omega (\gamma_2^2 + \delta^2)$. In the exact resonance ($\delta=0$), we have $\varepsilon'_{eff} = n_d^2$ and $\varepsilon''_{eff} = 3 \omega_L T_2 w_{eff}$, where $w_{eff}=(1 + |\Omega|^2/\Omega_{sat}^2)^{-1}$ is the sought-for effective population difference. Since $\varepsilon_{eff} = (n + i \kappa)^2$, we can easily connect $w_{eff}$ with the effective absorption coefficient $\kappa$, which, in turn, can be linked to the transmission as $T=\exp(-4 \pi \kappa L / \lambda)$. For the parameters used in our calculations, one should take $w_{eff} \approx 5.5 \cdot 10^{-4}$ to reach the transmission of $50 \%$. This value of effective population difference is close to zero (i.e. the medium is indeed saturated) and has the same order of magnitude as the stationary population difference obtained in our numerical calculations.

Note that in our estimation, we have neglected reflection which is indeed very low as seen in Fig. \ref{fig1}. This can be easily explained with a simple calculation of transmittion and reflection of light from a uniform layer with the effective permittivity having small imaginary part. Finally, we see from the expression $w_{eff}=(1 + |\Omega|^2/\Omega_{sat}^2)^{-1}$ that it should depend on the incident wave intensity: increasing intensity, we can make absorption smaller due to saturation. In other words, the low-intensity waves are almost entirely absorbed, whereas the high-intensity ones are mostly transmitted. The effect of disorder on the propagation time can be conveniently observed at the intermediate intensities, not very low and not very high (e.g., $\Omega_0=10 \gamma_2$ as in Fig. \ref{fig1}). 

The features discussed in this Section are also valid for two interacting wave fronts as shown in Supplementary Information.

\section{Adiabatic fronts}\label{adiab}

In this section, we consider the case of adiabatically switching field (\ref{waveform}), when $t_p \gg T_2$. In particular, we take the relaxation times $T_1=1$ ns and $T_2=0.1$ ps and the switching time $t_p=30 T_2$. The Lorentz frequency is $\omega^0_L = 10^{11}$ s$^{-1}$. The full thickness of the medium is $L=200 \lambda$, which is long enough for a kink to form and can be traversed after $Ln_d/c \sim 0.27 t_p$ in the case of dispersion-free medium. The final amplitude is $\Omega_0=0.3 \gamma_2$, so that there is no Rabi oscillations and the light-matter interaction is quasi-stationary in this case \cite{Novitsky2020}.

\begin{figure}[t!]
\centering \includegraphics[scale=1, clip=]{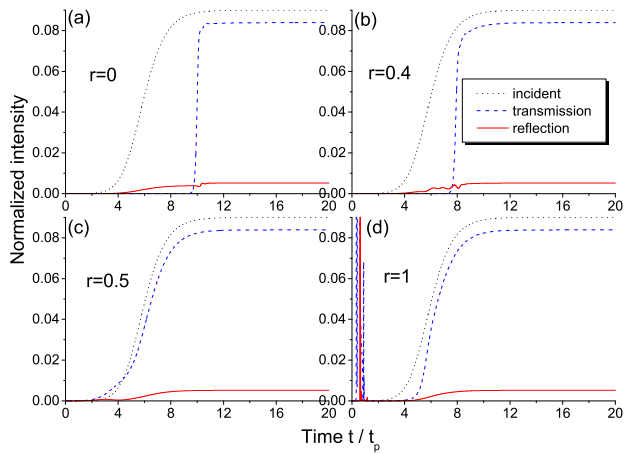}
\caption{\label{fig4} Intensity profiles for the transmitted and reflected light in the case of incident \textit{adiabatic} front. Different panels show the results for different disorder strengths $r$.}
\end{figure}

It is known that the adiabatically switching waveform (\ref{waveform}) undergoes self-steepening resulting in the kink (shock wave) formation after some distance passed through the resonantly absorbing medium \cite{Ponomarenko2010}. Such a kink is seen in Fig. \ref{fig4}(a) for the disorder strength $r=0$, when all the layers are the same absorbing medium. For increased disorder, we still obtain the kink at the exit [Fig. \ref{fig4}(b) at $r=0.4$], although the system is now non-uniform and contains both loss and gain layers. Note the increased speed of this kink. For $r=0.5$, when both loss and gain are equally probable, the kink formation is totally suppressed as shown in Fig. \ref{fig4}(c). Moreover, the transmitted intensity grows ahead of the incident intensity (at early times $t<4 t_p$). These features are due to the large number of gain layers providing the proper amplification of the signal and fast saturation of the medium, so that the wave can almost freely propagate through the structure in later times. For even stronger disorders, the portion of gain layers becomes so large that even tiny impinging radiation rapidly stimulates a powerful burst of energy as seen in Fig. \ref{fig4}(d) for $r=1$ (uniform gain medium). After this burst, the most part of the particles return to the ground state, so that the medium remains only weakly excited and gets saturated by the incident wave front. Although the kink is not formed in this case, the transmitted profile is closer to a kink than for $r=0.5$, manifesting a characteristic offset time between input and output signals.

Thus, disorder gives us an opportunity to control the shape of the output wave switching on and off kink formation. As shown in Supplementary Information, a similar effect disorder has on a pair of interacting adiabatic fronts.

\section{Conclusion}\label{concl}

In summary, we have proposed a novel approach for wave-front velocity and shape governing in the multilayer structure with disorderly distributed resonant loss and gain. This concept involves functionalities of both non-Hermitian system and disordered multilayer providing several new optical effects. The two types of fronts were considered -- the adiabatic (slowly switching) and non-adiabatic (rapidly switching) ones. These two cases have different dynamic features -- kinks and Rabi oscillations, respectively, -- which condition the propagation characteristics of the wave fronts. Introduction of disorder governed by external pumping results in a number of easily noticeable novel features, such as kink suppression and propagation time shortening. These disorder-induced effects together with the transmitted intensity determined by the medium saturation give us a tool to control transmission, shape and propagation time of the waves with the finite switching time. Such possibilities are extremely important for manipulation of light produced by realistic laser sources and can be used in optical switching and data processing among other applications.

\acknowledgements{The work was supported by the Belarusian Republican Foundation for Fundamental Research (Project No. F20R-158), the Russian Foundation for Basic Research (Projects No. 18-02-00414 and 20-52-00031), and Government of Russian Federation (Grant No. 08-08). Numerical simulations of the nonlinear interaction of light with resonant media have been supported by the Russian Science Foundation (Project No. 18-72-10127).}


\begin{thebibliography}{00}
\bibitem{Zyablovsky2014} A.~A.~Zyablovsky, A.~P.~Vinogradov, A.~A.~Pukhov, A.~V.~Dorofeenko, and A.~A.~Lisyansky,
$\mathcal{PT}$ symmetry in optics, {Phys. Usp.} \textbf{57}, 1063
(2014).

\bibitem{Feng2017} L.~Feng, R.~El-Ganainy, and L.~Ge,
Non-Hermitian photonics based on parity-time symmetry, {Nat.
Photon.} \textbf{11}, 752 (2017).

\bibitem{El-Ganainy2018} R.~El-Ganainy, K.~G.~Makris, M.~Khajavikhan, Z.~H.~Musslimani, S.~Rotter, and D.~N.~Christodoulides,
Non-Hermitian physics and $\mathcal{PT}$ symmetry, {Nat. Phys.}
\textbf{13}, 11 (2018).

\bibitem{Ozdemir2019} S.~K.~\"Ozdemir, S. Rotter, F. Nori, and L. Yang,
Parity-time symmetry and exceptional points in photonics, {Nat.
Mater.} \textbf{18}, 783 (2019).

\bibitem{Miri2019} M.-A. Miri and A. Alu, Exceptional points in optics and photonics, {Science} \textbf{363}, eaar7709 (2019).

\bibitem{Lin2011} Z.~Lin, H.~Ramezani, T.~Eichelkraut, T.~Kottos, H.~Cao, and D.~N.~Christodoulides,
Unidirectional invisibility induced by $\mathcal{PT}$-symmetric
periodic structures, {Phys. Rev. Lett.} \textbf{106}, 213901 (2011).

\bibitem{Feng2013} L. Feng, Y.-L. Xu, W. S. Fegadolli, M.-H. Lu, J. E. B. Oliveira, V.
R. Almeida, Y.-F. Chen, and A. Scherer, Experimental demonstration
of a unidirectional reflectionless paritytime metamaterial at
optical frequencies, \textit{Nat. Mater.} \textbf{12}, 108 (2013).

\bibitem{Chen2017} W.~Chen, S.~K.~\"{O}zdemir, G.~Zhao, J.~Wiersig, and L.~Yang,
Exceptional points enhance sensing in an optical microcavity,
{Nature (London)} \textbf{548}, 192 (2017).

\bibitem{Hodaei2017} H.~Hodaei, A.~U.~Hassan, S.~Wittek, H.~Garcia-Gracia,
R.~El-Ganainy, D.~N.~Christodoulides, and M.~Khajavikhan, Enhanced
sensitivity at higher-order exceptional points, {Nature (London)}
\textbf{548}, 187 (2017).

\bibitem{Hokmabadi2019} M.~P.~Hokmabadi, A.~Schumer, D.~N.~Christodoulides, and M.~Khajavikhan, Non-Hermitian ring laser gyroscopes with enhanced Sagnac sensitivity, {Nature (London)} \textbf{576}, 70 (2019).

\bibitem{Lai2019} Y.-H.~Lai, Y.-K.~Lu, M.-G.~Suh, Z.~Yuan, and K.~Vahala, Observation of the exceptional-point-enhanced Sagnac effect, {Nature (London)} \textbf{576}, 65 (2019).

\bibitem{Peng2014} B.~Peng, S.K.~\"Ozdemir, S.~Rotter, H.~Yilmaz, M.~Liertzer, F.~Monifi, C.~M.~Bender, F.~Nori, and L.~Yang, Loss-induced suppression and revival of lasing, {Science} {\bf346}, 328 (2014).

\bibitem{Novitsky2018} D.~V.~Novitsky, A.~Karabchevsky, A.~V.~Lavrinenko, A.~S.~Shalin, and A.~V.~Novitsky, $\mathcal{PT}$ symmetry breaking in multilayers with resonant loss and gain locks light propagation direction {Phys. Rev. B} {\bf 98}, 125102 (2018).

\bibitem{Feng2014-2} L.~Feng, Z.~J.~Wong, R.-M.~Ma, Y.~Wang, and X.~Zhang,
Single-mode laser by parity-time symmetry breaking, {Science}
\textbf{346}, 972 (2014).

\bibitem{Hodaei2014} H.~Hodaei, M.-A.~Miri, M.~Heinrich, D.~N.~Christodoulides, and M.~Khajavikhan, Parity-time-symmetric microring lasers, {Science} \textbf{346}, 975 (2014).

\bibitem{Gu2016} Z.~Gu, N.~Zhang, Q.~Lyu, M.~Li, S.~Xiao, and Q.~Song,
Experimental demonstration of $\mathcal{PT}$-symmetric stripe lasers, {Laser Photon. Rev.} \textbf{10}, 588 (2016).

\bibitem{Zhang2020} Z.~Zhang, X.~Qiao, B.~Midya, K.~Liu, J.~Sun, T.~Wu, W.~Liu, R.~Agarwal, J.~M.~Jornet, S.~Longhi, N.~M.~Litchinitser, and L.~Feng, Tunable topological charge vortex microlaser, {Science} \textbf{368}, 760 (2020).

\bibitem{Longhi2010} S.~Longhi,
$\mathcal{PT}$-symmetric laser absorber, {Phys. Rev. A} \textbf{82}, 031801(R) (2010).

\bibitem{Chong2011} Y.~D.~Chong, L.~Ge, and A.~D.~Stone,
Symmetry Breaking and Laser-Absorber Modes in Optical Scattering
Systems, {Phys. Rev. Lett.} \textbf{106}, 093902 (2011).

\bibitem{Wong2016} Z.~J.~Wong, Y.-L.~Xu, J.~Kim, K.~O'Brien, Y.~Wang, L.~Feng, and X.~Zhang,
Lasing and anti-lasing in a single cavity, {Nat. Photon.}
\textbf{10}, 796 (2016).

\bibitem{Novitsky2019a} D.~V.~Novitsky, CPA-laser effect and exceptional points in $\mathcal{PT}$-symmetric multilayer structures {J. Opt.} {\bf 21}, 085101 (2019).

\bibitem{Leykam2017} D.~Leykam, K.~Y.~Bliokh, C.~Huang, Y.~D.~Chong, and F.~Nori, Edge Modes, Degeneracies, and Topological Numbers in Non-Hermitian Systems, {Phys. Rev. Lett.} {\bf118}, 040401 (2017).

\bibitem{Takata2018} K.~Takata and M.~Notomi, Photonic Topological Insulating Phase Induced Solely by Gain and Loss, {Phys. Rev. Lett.} {\bf121}, 213902 (2018).

\bibitem{Ni2018} X.~Ni, D.~Smirnova, A.~Poddubny, D.~Leykam, Y.~Chong, and A.~B.~Khanikaev, $\mathcal{PT}$ phase transitions of edge states at $\mathcal{PT}$ symmetric interfaces in non-Hermitian topological insulators, {Phys. Rev. B} {\bf98}, 165129 (2018).

\bibitem{Zhao2019} H.~Zhao, X.~Qiao, T.~Wu, B.~Midya, S.~Longhi, and L.~Feng, Non-Hermitian topological light steering, {Science} {\bf365}, 1163 (2019).

\bibitem{Wiersma2013} D.~S.~Wiersma, Disordered photonics, {Nat. Photon.} {\bf7}, 188 (2013).

\bibitem{Zhigunov2018} D.~Zhigunov, A.~B.~Evlyukhin, A.~S.~Shalin, U.~Zywietz, and B.~N.~Chichkov, Femtosecond laser printing of single Ge and SiGe nanoparticles with electric and magnetic optical resonances, {ACS Photon.} {\bf5}, 977 (2018).

\bibitem{Baryshnikova2018} K.~Baryshnikova, D.~Filonov, C.~Simovski, A.~Evlyukhin, A.~Kadochkin, E.~Nenasheva, P.~Ginzburg, and A.~S.~Shalin, Giant magnetoelectric field separation via anapole-type states in high-index dielectric structures, {Phys. Rev. B} {\bf98}, 165419 (2018).

\bibitem{Ivinskaya2018} A.~Ivinskaya, N.~Kostina, A.~Proskurin, M.~I.~Petrov, A.~A.~Bogdanov, S.~Sukhov, A.~V.~Krasavin, A.~Karabchevsky, A.~S.~Shalin, and P.~Ginzburg, Optomechanical Manipulation with Hyperbolic Metasurfaces, {ACS Photon.} {\bf5}, 4371 (2018).

\bibitem{Segev2013} M.~Segev, Y.~Silberberg, and D.~N.~Christodoulides, Anderson localization of light, {Nat. Photon.} {\bf7}, 197 (2013).

\bibitem{Gredeskul2012} S.~A.~Gredeskul, Yu.~S.~Kivshar, A.~A.~Asatryan, K.~Y.~Bliokh, Yu.~P.~Bliokh, V.~D.~Freilikher, and I.~V.~Shadrivov, Anderson localization in metamaterials and other complex media, {Low Temp. Phys.} {\bf38}, 570 (2012).

\bibitem{Sheinfux2017} H.~H.~Sheinfux, Y.~Lumer, G.~Ankonina, A.~Z.~Genack, G.~Bartal, and M.~Segev, Observation of Anderson localization in disordered nanophotonic structures, {Science} {\bf356}, 953 (2017).

\bibitem{Chabanov2003} A.~A.~Chabanov, Z.~Q.~Zhang, and A.~Z.~Genack, Breakdown of Diffusion in Dynamics of Extended Waves in Mesoscopic Media, {Phys. Rev. Lett.} {\bf90}, 203903 (2003).

\bibitem{Sarma2014} R.~Sarma, T.~Golubev, A.~Yamilov, and H.~Cao, Control of light diffusion in a disordered photonic waveguide, {Appl. Phys. Lett.} {\bf105}, 041104 (2014).

\bibitem{Naraghi2016} R.~R.~Naraghi and A.~Dogariu, Phase Transitions in Diffusion of Light, {Phys. Rev. Lett.} {\bf117}, 263901 (2016).

\bibitem{Conti2007} C.~Conti, L.~Angelani, and G.~Ruocco, Light diffusion and localization in three-dimensional nonlinear disordered media, {Phys. Rev. A} {\bf75}, 033812 (2007).

\bibitem{Lahini2008} Y.~Lahini, A.~Avidan, F.~Pozzi, M.~Sorel, R.~Morandotti, D.~N.~Christodoulides, and Y.~Silberberg, Anderson Localization and Nonlinearity in One-Dimensional Disordered Photonic Lattices, {Phys. Rev. Lett.} {\bf100}, 013906 (2008).

\bibitem{Sharabi2018} Y.~Sharabi, H.~H.~Sheinfux, Y.~Sagi, G.~Eisenstein, and M.~Segev, Self-Induced Diffusion in Disordered Nonlinear Photonic Media, {Phys. Rev. Lett.} {\bf121}, 233901 (2018).

\bibitem{Makris2015} K.~G.~Makris, Z.~H.~Musslimani, D.~N.~Christodoulides, and S.~Rotter, Constant-intensity waves and their modulation instability in non-Hermitian potentials, {Nat. Commun.} {\bf6}, 7257 (2015).

\bibitem{Makris2015} K.~G.~Makris, Z.~H.~Musslimani, D.~N.~Christodoulides, and S.~Rotter, Constant-intensity waves and their modulation instability in non-Hermitian potentials, {Nat. Commun.} {\bf6}, 7257 (2015).

\bibitem{Makris2017} K.~G.~Makris, A.~Brandst\"{o}tter, P.~Ambichl, Z.~H.~Musslimani, and S.~Rotter, Wave propagation through disordered media without backscattering and intensity variations, {Light Sci. Appl.} {\bf6}, e17035 (2017).

\bibitem{Brandstotter2019} A.~Brandst\"{o}tter, K.~G.~Makris, and S.~Rotter, Scattering-free pulse propagation through invisible non-Hermitian media, {Phys. Rev. B} {\bf99}, 115402 (2019).

\bibitem{Tzortzakakis2020} A.~F.~Tzortzakakis, K.~G.~Makris, S.~Rotter, and E.~N.~Economou, Shape-preserving beam transmission through non-Hermitian disordered lattices, arXiv:2005.06414 (2020).

\bibitem{Makris2020} K.~G.~Makris, I.~Kre\v{s}i\'{c}, A.~Brandst\"{o}tter, and S.~Rotter, Scattering-free channels of invisibility across non-Hermitian media, {Optica} {\bf7}, 619 (2020).

\bibitem{Hamazaki2019} R.~Hamazaki, K.~Kawabata, and M.~Ueda, Non-Hermitian Many-Body Localization, {Phys. Rev. Lett.} {\bf123}, 090603 (2019).

\bibitem{Tzortzakakis2020a} A.~F.~Tzortzakakis, K.~G.~Makris, and E.~N.~Economou, Non-Hermitian disorder in two-dimensional optical lattices, {Phys. Rev. B} {\bf101}, 014202 (2020).

\bibitem{Huang2020} Y. Huang and B.~I.~Shklovskii, Anderson transition in three-dimensional systems with non-Hermitian disorder, {Phys. Rev. B} {\bf101}, 014204 (2020).

\bibitem{Noginov2005} M.~A.~Noginov, \textit{Solid-State Random Lasers} (Springer, New York, 2005).

\bibitem{Turitsyn2010} S.~K.~Turitsyn, S.~A.~Babin, A.~E.~El-Taher, P.~Harper, D.~V.~Churkin, S.~I.~Kablukov, J.~D.~Ania-Casta\~{n}\'{o}n, V.~Karalekas, and E.~V.~Podivilov, Random distributed feedback ﬁbre laser, {Nat. Photon.} {\bf4}, 231 (2010).

\bibitem{Liu2014} J.~Liu, P.~D.~Garcia, S.~Ek, N.~Gregersen, T.~Suhr, M.~Schubert, J.~M{\o}rk, S.~Stobbe, and P.~Lodahl, Random nanolasing in the Anderson localized regime, {Nat. Nanotech.} {\bf9}, 285 (2014).

\bibitem{Abaie2017} B.~Abaie, E.~Mobini, S.~Karbasi, T.~Hawkins, J.~Ballato, and A.~Mafi, Random lasing in an Anderson localizing optical ﬁber, {Light Sci. Appl.} {\bf6}, e17041 (2017).

\bibitem{Novitsky2019} D.~V.~Novitsky, D.~Redka, and A.~S.~Shalin, Diﬀerent Regimes of Ultrashort Pulse Propagation in Disordered Layered Media with Resonant Loss and Gain, {Ann. Phys. (Berlin)} {\bf 531}, 1900080 (2019).

\bibitem{Allen} L.~Allen and J.~H.~Eberly, \textit{Optical Resonance and Two-Level Atoms} (Wiley, New York, 1975).

\bibitem{Crenshaw2008} M.~E.~Crenshaw, Comparison of quantum and classical local-ﬁeld effects on two-level atoms in a dielectric, {Phys. Rev. A} {\bf78}, 053827 (2008).

\bibitem{Novitsky2011} D.~V.~Novitsky, Femtosecond pulses in a dense two-level medium: Spectral transformations, transient processes, and collisional dynamics, {Phys. Rev. A} {\bf84}, 013817 (2011).

\bibitem{Crenshaw1996} M.~E.~Crenshaw, Quasiadiabatic approximation for a dense collection of two-level atoms, {Phys. Rev. A} {\bf54}, 3559 (1996).

\bibitem{Novitsky2009} D.~V.~Novitsky, Compression of an intensive light pulse in photonic-band-gap structures with a dense resonant medium, {Phys. Rev. A} {\bf79}, 023828 (2009).

\bibitem{Bachelard2014} N.~Bachelard, S.~Gigan, X.~Noblin, P.~Sebbah, Adaptive pumping for spectral control of random lasers, {Nat. Phys.} \textbf{10}, 426 (2014).

\bibitem{Lee2019} M.~Lee, S.~Callard, C.~Seassal, and H.~Jeon, Taming of random lasers, {Nat. Photon.} \textbf{13}, 445 (2019).

\bibitem{Novitsky2020} D.~V.~Novitsky and A.~S.~Shalin, Kink-based mirrorless quasi-bistability in resonantly absorbing media, {Opt. Lett.} {\bf 45}, 137 (2020).

\bibitem{Ponomarenko2010} S.~A.~Ponomarenko and S.~Haghgoo, Self-similarity and optical kinks in resonant nonlinear media, {Phys. Rev. A} {\bf82}, 051801(R) (2010).

\bibitem{Novitsky2017} D.~V.~Novitsky, V.~R.~Tuz, S.~L.~Prosvirnin, A.~V.~Lavrinenko, and A.~V.~Novitsky, Transmission enhancement in loss-gain multilayers by resonant suppression of reﬂection, {Phys. Rev. B} {\bf 96}, 235129 (2017).

\bibitem{Novitsky2017a} D.~V.~Novitsky, Optical kinks and kink-kink and kink-pulse interactions in resonant two-level media, {Phys. Rev. A} {\bf 95}, 053846 (2017).
\end{thebibliography}
\end{document}


\title{Supplementary Information for \\ Controlling wave-front shape and propagation time with tunable disordered non-Hermitian multilayers}

\author{Denis~V.~Novitsky$^{1}$}
\author{Dmitry~Lyakhov$^{2}$}
\author{Dominik~Michels$^{2}$}
\author{Dmitrii Redka$^3$}
\author{Alexander~A.~Pavlov$^{4}$}
\author{Alexander~S.~Shalin$^{5}$}

\affiliation{$^1$B.~I.~Stepanov Institute of Physics, National
Academy of Sciences of Belarus, Nezavisimosti Avenue 68, 220072
Minsk, Belarus \\ $^2$Visual Computing Center, King Abdullah University of Science and Technology, Thuwal 23955-6900, Kingdom of Saudi Arabia\\ $^3$ Saint Petersburg, Electrotechnical University “LETI” (ETU),  Prof. Popova Street 5, 197376 St. Petersburg, Russia \\ $^4$ Institute of Nanotechnology of Microelectronics of the Russian Academy of Sciences, Leninsky Prospekt 32A, 119991 Moscow, Russia \\ $^5$ITMO University, Kronverksky Prospekt 49,
197101 St. Petersburg, Russia}

\maketitle

We have seen that changing the disorder strength allows to control propagation time of the wave front through the loss-gain multilayer. Another approach to such control is to use two waves in the counter-propagating fashion: the propagation time of a wave depends on the intensity of another one \cite{Novitsky2017a}. Here we study the combination of both approaches considering interaction of wave fronts in the presence of disorder.

\begin{figure}[b!]
\centering \includegraphics[scale=1, clip=]{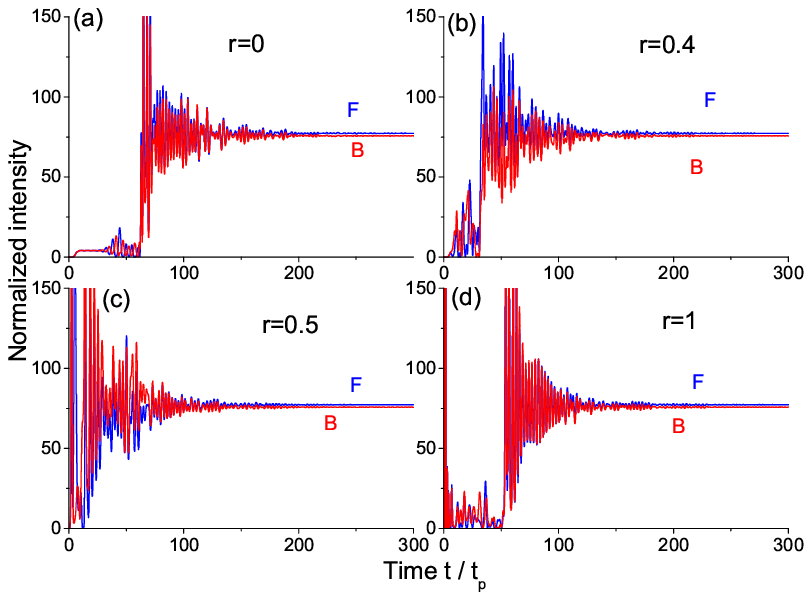}
\caption{\label{fig5} Intensity profiles for the forward (F) and backward (B) transmitted waves in the case of incident non-adiabatic fronts. Different panels show the results for different disorder strengths $r$.}
\end{figure}

We start with the case of two identical non-adiabatic fronts with one of them being ``forward-propagating'' and another being ``backward-propagating''. The parameters of calculation are the same as in corresponding section of the main text, the amplitude of the waves being $\Omega_0=10 \gamma_2$. The results of calculation for different $r$ are shown in Fig. \ref{fig5}. One can see that the time needed for the wave front to traverse the system and to establish the steady-state response becomes shorter in comparison to the single-front case (see Fig. 2 of the main text). This is in full accordance with the previous report \cite{Novitsky2017a}. On the other hand, at first, increase of disorder results in shortening propagation time (compare the cases of $r=0$ and $r=0.5$ in Fig. \ref{fig5}). Then, it grows again and at $r=1$ is practically the same as at $r=0$, in agreement with the regularities discussed for single fronts.

\begin{figure}[t!]
\centering \includegraphics[scale=1, clip=]{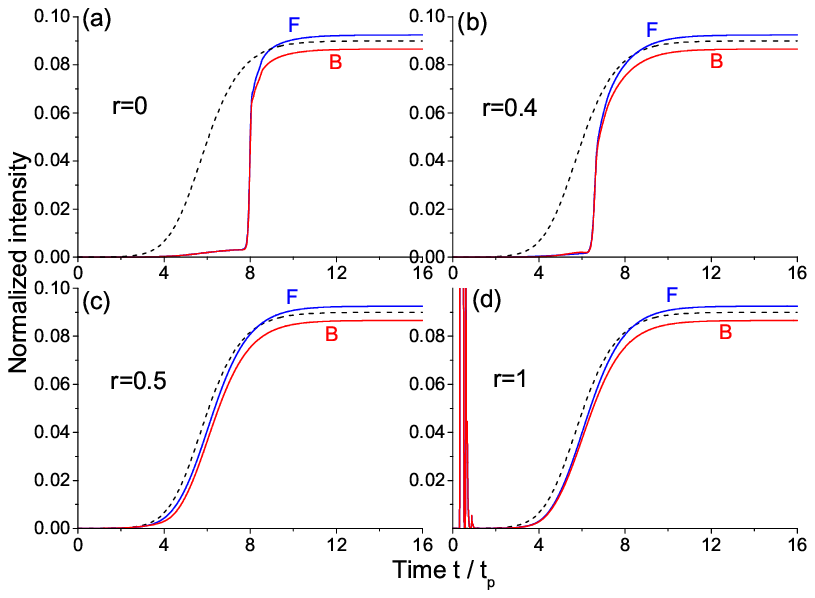}
\caption{\label{fig6} Intensity profiles for the forward (F) and backward (B) transmitted waves in the case of incident adiabatic fronts. Different panels show the results for different disorder strengths $r$.}
\end{figure}

The similar features occur for the adiabatic fronts as shown in Fig. \ref{fig6} for the same parameters as in the main text. The interacting kinks seen in panels (a) and (b) pass through the medium for the shorter time than the single kink. The only peculiarity is the low pedestal preceding the kink itself. This can be attributed to partial reflection of the counter-propagating wave. For larger disorder strengths ($r \gtrsim 0.5$), formation of the kinks is suppressed as well as for the single wave front (compare with Fig. 5 in the main text).

